\documentclass[3p]{elsarticle}
\pdfoutput=1

\usepackage{xspace}

\newcommand{\approach}{SATDBailiff\xspace}

\usepackage{pgf-pie}
\usepackage{adjustbox}
\usepackage{dirtytalk}
\usepackage{amsmath,amssymb,amsfonts}
\usepackage{algorithmic}
\usepackage{graphicx}
\usepackage{textcomp}
\usepackage{xcolor}
\usepackage{tikz}
\usetikzlibrary{backgrounds}
\usepackage{verbatim}
\usepackage{ulem}
\usepackage{fancyvrb}
\usepackage{float}

\def\BibTeX{{\rm B\kern-.05em{\sc i\kern-.025em b}\kern-.08em
    T\kern-.1667em\lower.7ex\hbox{E}\kern-.125emX}}

\usepackage{soul}
\usepackage{listings}
\colorlet{soulred}{red!30}
\colorlet{soulgreen}{green!30}
\colorlet{soulredchange}{red!50}
\colorlet{soulgreenchange}{green!50}
\DeclareRobustCommand{\hlremove}[1]{{\sethlcolor{soulred}\hl{#1}}}
\DeclareRobustCommand{\hladd}[1]{{\sethlcolor{soulgreen}\hl{#1}}}
\DeclareRobustCommand{\hlremovechange}[1]{{\sethlcolor{soulredchange}\hl{#1}}}
\DeclareRobustCommand{\hladdchange}[1]{{\sethlcolor{soulgreenchange}\hl{#1}}}
\newcommand*\circled[1]{\tikz[baseline=(char.base)]{
            \node[shape=circle,draw,inner sep=2pt] (char) {#1};}}
    
\usepackage{amssymb}
 
\usepackage[figuresright]{rotating}
\usepackage{subcaption}

\usepackage[english]{babel}     

\usepackage{amsmath}            
\usepackage{epstopdf}           
\usepackage{flushend}           
\usepackage{hyperref}           

\usepackage[most]{tcolorbox}   

\usepackage{textcomp}  
\newcommand{\eman}[1]{\textcolor{red}{{\it [Eman says: #1]}}}

\usepackage{mathtools}
\usepackage{amsmath}

\usepackage{tabularx}
\usepackage{pgfplots}
\pgfplotsset{width=7cm,compat=1.8,tick label style={font=\small}}
\usepackage{pgfplotstable}
\usepackage{sfmath}
\usepackage{array, booktabs, makecell}
\usepackage{color}
\usepackage{csquotes}
\usepackage{url}
\usepackage{comment}
\usepackage{tikz}
\usepackage{balance}
\usepackage{listings}
\usepackage{breqn}
\usepackage{booktabs} 
\usepackage{soul} 
\usetikzlibrary{fit} 
\usetikzlibrary{positioning}
\usepackage{courier}

\usepackage{multirow}

\usepackage{array}
\usepackage{makecell}

\usepackage{array,booktabs,ragged2e}
\newcolumntype{R}[1]{>{\RaggedLeft\arraybackslash}p{#1}}

\usepackage{svg}

\definecolor{findOptimalPartition}{HTML}{D7191C}
\definecolor{storeClusterComponent}{HTML}{FDAE61}
\definecolor{dbscan}{HTML}{ABDDA4}
\definecolor{constructCluster}{HTML}{2B83BA}

\begin{document}
\selectlanguage{english}
\begin{frontmatter}

\title{\approach  - Mining and Tracking Self-Admitted Technical Debt}

\author{Eman Abdullah AlOmar, Ben Christians, Mihal Busho, Ahmed Hamad AlKhalid, Ali Ouni, Christian Newman, Mohamed Wiem Mkaouer\corref{cor1}}
\ead{\{eaa6167,bbc7909,mb5185,aa5130,cdnvse,mwmvse\}@rit.edu,ali.ouni@etsmtl.ca}

\address{Rochester Institute of Technology, NY, USA \\ ETS Montreal, University of Quebec, Canada}


\begin{abstract}
Self-Admitted Technical Debt (SATD) is a metaphorical concept to describe the self-documented addition of technical debt to a software project in the form of source code comments.
SATD can linger in projects and degrade source-code quality, but it can also be more visible than unintentionally added or undocumented technical debt.
Understanding the implications of adding SATD to a software project is important because developers can benefit from a better understanding of the quality trade-offs they are making.
However, empirical studies, analyzing the survivability and removal of SATD comments, are challenged by potential code changes or SATD comment updates that may interfere with properly tracking their appearance, existence, and removal. 
In this paper, we propose \approach, a tool that uses an existing state-of-the-art SATD detection tool, to identify SATD in method comments, then properly track their \textit{lifespan}. \approach is given as input links to open source projects, and its output is a list of all identified SATDs, and for each detected SATD, \approach reports all its associated changes, including any updates to its text, all the way to reporting its removal. The goal of \approach is to aid researchers and practitioners in better tracking SATDs instances, and providing them with a reliable tool that can be easily extended.
\approach was validated using a dataset of previously detected and manually validated SATD instances. 
\approach is publicly available as an open source, along with the manual analysis of SATD instances associated with its validation, on the project website\footnote{\url{https://smilevo.github.io/self-affirmed-refactoring/SCP20_index.html}}.

\end{abstract}


\end{frontmatter}

\section{Introduction}
Technical debt (TD) is a metaphor that describes taking shortcuts in software development that will require additional time to fix (or payback) in the future \cite{Kruchten}. Technical debt commonly occurs when developers conclude development of a section of code before it is complete and optimized \cite{cunningham1992wycash}. This is done with an understanding that this creates a \textit{debt} that will need additional time and effort to be managed later on. Developers tend to understand \textit{some} quality-related implications of adding this debt to their projects, which can be seen unmistakably when looking specifically at \textit{Self-Admitted Technical Debt} (SATD) \cite{potdar2014exploratory}. 

Self-Admitted Technical Debt is a candid form of technical debt in which the contributor of the debt self-documents the location of the debt. This admission is typically accompanied by a description of a known or potential defect or a statement detailing what remaining work must be done. Well-known and frequently used examples of SATD include comments beginning with \textit{TODO}, \textit{FIXME}, \textit{BUG}, \textit{XXX}, or \textit{HACK}. SATD can also take other forms of more complex language void of any of the previously mentioned keywords. Any comment detailing a \textit{not-quite-right} implementation present in the surrounding code can be classified as SATD.

Modern Integrated Development Environments (IDEs) have begun recognizing the utility of SATD. It is common for them to highlight comments containing the aforementioned keywords, or for them to add SATD to a project when automatically generating unimplemented stubbed functionality to be implemented manually at a later time. Developers and IDEs both contribute SATD with the common assumption that including a self-admission will make their technical debt easier to pay back, or at least reduce the likelihood of it being forgotten. The effectiveness of this strategy needs to be brought into question, as it may have significant impacts on development practices. Understanding the implications of this assumption is vital to assure high-quality performance in software development teams.

Since SATD is a written testament of an existing negative manifestation in the source code, several studies have observed the removal of SATD instances to better understand how developers manage and resolve TD. Detecting the removal of TD is of interest to both researchers and practitioners, as, in addition to indicating the disappearance of a problem, it indicates the changes containing the fix to that TD. Such TD fixes are important to locate, since they can be valuable when taking corrective actions against similar TDs. Several recent studies have been focused on accurately identifying the removal of SATD. For instance, Bavota and Russo \cite{bavota2016large} have shown that up to 57\% of SATD is addressed, and those instances are typically addressed by the same developer who initially contributed to the SATD. However, the identification of SATD removal is complex when taking into account common occurrences when code changes overshadow SATD removals \cite{wehaibi2016examining}, such as the renaming of code elements containing the instance or the accidental deletion of containing source files. Another challenge that threatens the correctness of empirical studies related to SATD is the changes that may interfere with properly tracking the survivability of SATDs. For example, if the class containing the SATD gets renamed or moved, without properly handling such refactorings, one SATD may look like it was deleted, and another one appeared, while it is the same SATD. Similarly, if the SATD text gets updated, without properly handling such change, this can be detected by a removal of one SATD instance, and the appearance of another one.  While not all accidental \textit{disappearances} of SATD comments imply the correction of any associated technical debt, there is a need for a tool that can reliably track the appearance and removal of SATD comments, along with any potential changes associated with the text of the SATD. This effort would provide valuable support for existing studies by clearly capturing these removals without the need to manually validate them.

To address the above-mentioned challenge, in this tool paper, we propose SATDBailiff, a tool based on the existing classification model, called SATD Detector \cite{liu2018satd}. This tool is able to (i) mine, identify, and track the additions, removal, and changes to SATD comments, while providing an overview of their \textit{lifespan} in the project, (ii) detect all textual changes associated with the SATD, detect all changes associated with the class containing the SATD (moving, renaming) that it underwent throughout the later commits up until the commit in which each instance was removed, if applicable, and (iii) allow the integration of SATD detection and classification tools, and (iv) allow the input of software repository link, and the output of all commits associated with additions, changes, and deletions of identified SATDs.




\approach was validated using a dataset of previously detected and manually validated SATD instances \cite{Maldonado2017-1}. \approach is challenged in identifying and tracking those instances throughout the evolution of five long-lived open-source projects from different application domains, namely \textit{Gerrit}, \textit{Camel}, \textit{Hadoop}, \textit{Log4j}, and \textit{Tomcat}. The detail of the studied projects is summarized in Table~\ref{tab:Studied Projects}. We manually analyzed the ability of \approach in correctly identifying SATD changes and removals (SATD additions were provided by the dataset). Results show that \approach is efficient by averaging an accuracy score of \textit{0.97}  when tracking SATD instances from their appearance in the project until their disappearance.

\textbf{Tool, documentation, docker and demo video.} \approach is publicly available as an open source tool\footnote{\url{https://github.com/smilevo/SATDBailiff}}, along with a continuous integration feature with a docker and a demo video. The raw data and the manual analysis of SATD instances are also available on the project website.\footnote{\url{https://smilevo.github.io/self-affirmed-refactoring/SCP20_index.html}}

The rest of the paper is organized as follows. Section \ref{section:background} gives an overview of the necessary information related to SATD and summarizes the related work. Section \ref{section:approach} describes our approach, \approach. Section \ref{section:usage} describes how \approach  can be used in practice while Section \ref{section:applicability} shows the applicability of the tool. Section \ref{section:validation} details the results of our experiments to evaluate \approach. In Section \ref{section:limitations}, we report the tool's limitation. Section \ref{section:threats} discusses the known threats to validity, while Section \ref{section:conclusion} draws our conclusions and future investigations.

\begin{table}
    \centering
    \caption{Details of the studied 5 projects}
    \begin{tabular}{ |p{4em}|R{5em}|R{5.2em}|R{4.8em}|R{4.8em}|   }
     \hline
     \textbf{Project} & \textbf{\# Java files} & \textbf{\# SLOC} &  \textbf{File versions}&  \textbf{\# Contributors}\\
     \hline 
     Camel\footnote{\url{https://github.com/apache/camel}}          & 15,091  & 800,488  &  254,920 & 289\\
     Gerrit\footnote{\url{https://github.com/GerritCodeReview/gerrit}}         & 3,059  & 222,476  & 53,298 &  270\\
     Hadoop\footnote{\url{https://github.com/apache/hadoop}}         & 8,466  &  996,877  & 79,232 & 160 \\
     Log4j\footnote{\url{https://github.com/apache/log4j}}          & 1,112   & 30,287   &12,609 & 35 \\ 
     Tomcat\footnote{\url{https://github.com/apache/Tomcat }}         & 3,187  &  297,828  &  46,716 & 32 \\
    
     \hline
    \end{tabular}
    
    \label{tab:Studied Projects}
\end{table}
\section{Background \& Related Work}
\label{section:background}

This paper focuses on mining and tracking SATD instances from Git repositories. Thus, in this section, we are interested in discussing related work on SATD. A 
summary these state-of-the-art studies is depicted in Table \ref{Table:satd in Related Work}.

The investigation of Self-Admitted Technical Debt began to gain traction in 2014 with the study of Potdar and Shihab \cite{potdar2014exploratory}. Initial approaches to classifying source comments as SATD involved intensive manual efforts. Potdar and Shihab manually classified 101,762 Java code comments and generated a string matching heuristic based on 62 commonly occurring comment patterns. This heuristic inspired Maldonado and Shihab \cite{Maldonado2015} to apply this classification to ten projects by manually analyzing 33 K code comments. They indicated that SATD items point to five types of debt: design, requirement, defect, test, and documentation. Their findings show that design debt is the most common type of debt.  Later, Bavota and Russo \cite{bavota2016large} replicated the work of Potdar and Shihab and carried out an empirical study across 159 projects to explore the diffusion and evolution of SATD and its impact on software quality. Moreover, Wehaibi \textit{et al.} \cite{wehaibi2016examining} investigated the relationship between SATD and quality. They show that technical debt increases the system's difficulty and may lead to complex software changes.

Maldonado \textit{et al.} \cite{Maldonado2017-1} expanded the classification approach to use Natural Language Processing (NLP) to identify design and requirement debts from ten open source projects. They also reported the top-10 words appearing within design and requirement SATD-based comments.  Despite the increased performance of classification models, little work has been done to develop an empirical understanding of SATD in Java projects. One of the empirical studies has been conducted by Maldonado \textit{et al.} \cite{Maldonado2017-2} to analyze SATD comments removal by looking at the change history of five large open source projects.  Results of the study indicate that there is a high percentage of SATD comments removed and their survivability varies by project. The quality of this dataset has been brought into question by Zampetti \textit{et al.} \cite{Zampetti}, who manually altered and improved the dataset in an effort to improve its quality. While this filtered dataset is regarded to be high quality, the filtering process removed a significant number of entries, there does not exist a means to expand its size. In their in-depth investigation of SATD removal, they found that between 20\% and 50\% of SATD comments are removed when either the whole class or method is removed. Another study examined the relationship between refactoring and technical debt. Lammarino \textit{et al.} \cite{iammarino2019self}  particularly studied the co-occurrence of refactorings and SATD removals. The authors show that refactorings are more likely to co-occur with SATD removals than with other commits. Huang \textit{et al.} \cite{Huang} developed a new SATD classification model using text mining. They built a composite classifier that combined multiple classifiers from eight different source projects which improved the F1-score of classification over Maldonado \textit{et al.} by 27.95\%. Liu \textit{et al.} \cite{liu2018satd} proposed an Eclipse plugin and Java library tool called SATD detector to automatically detect SATD using text mining and highlight the detected comments in an integrated development environment (IDE).

More recently, Farias \textit{et al.} \cite{de2020identifying} improved a set of contextualized patterns or SATD identification vocabulary built to detect SATD by carrying out three empirical studies. With regard to the technical debt items detection, their results show that more than half of the new patterns were considered decisive or very decisive. In a different work, Zampetti \textit{et al.} \cite{zampetti2020automatically} developed SAtd Removal using DEep LEarning (SARDELE) that highlighted developers' need to cope with SATD removal.  Their tool is capable of recommending six SATD removal strategies (e.g., telling that a more complex change is needed). Their evaluation reveals that SARDELE is able to predict the type of change with an average precision of 55\% and recall of 57\%. In another study, Xavier \textit{et al.} \cite{xavier2020beyond} mined SATD in issue tracker systems.
They studied a sample of 286 SATD instances collected from five open source projects. Although  SATD instances are not more complex in terms of code churn, their findings show that SATD instances take more time to be closed.

There is now an opportunity to take advantage of the improved detection tool by Huang \textit{et al.} \cite{Huang} to enhance research efforts with a highly accurate, large scale empirical history of SATD instances in Java projects previously unavailable. This can be accomplished alongside fixing some of the data quality issues noted with Maldonado \textit{et al.}'s \cite{Maldonado2017-2} empirical study. This study will aim to package these improvements and model in a tool will allow further efforts to expand past these seven previously available software projects in terms of size and quality. In addition to the publication of this tool, an empirical history
of SATD instances in 30 open source software projects will be made available as produced by  \approach.

\begin{table*}
  \centering
	 \caption{A summary of the literature on SATD}
	 \label{Table:satd in Related Work}
\begin{adjustbox}{width=1.0\textwidth,center}
\begin{tabular}{|l|l|l|l|l|l|l|}\hline
\bfseries Study & \bfseries Year & \bfseries Focus & \bfseries Detection technique & \bfseries SATD tool & \bfseries SATD type & \bfseries Project size  \\ \hline
Potdar \& Shihab \cite{potdar2014exploratory} & 2014 & SATD identification & Manual analysis & N/A &  Not mentioned & 4   \\ \hline
Maldonado \& Shihab \cite{Maldonado2015} & 2015 & SATD detection & Manual analysis & N/A & design/requirement/defect/test & 10  \\ 
& & & & & documentation & \\ \hline
Bavota \& Russo \cite{bavota2016large} & 2016 & SATD identification & Mining-based technique & N/A & design/requirement/code/test & 159 \\ 
& & & & & architecture/defect/people/build & \\
& & & & & documentation & \\ \hline 
Wehaibi \textit{et al.} \cite{wehaibi2016examining} & 2016 & Impact of SATD on quality & Defect-based measurement & N/A & Not mentioned & 5  \\ \hline
Maldonado \textit{et al.} \cite{Maldonado2017-1} & 2017 & SATD automatic classification  & NLP-based technique & Maximum entropy classifier & design / requirement & 10  \\ \hline
Maldonado \textit{et al.} \cite{Maldonado2017-2} & 2017 & SATD removal & NLP-based technique & Maldonado \textit{et al.}'s tool & Not mentioned & 5  \\ \hline
Zampetti \textit{et al.} \cite{Zampetti} & 2018 & SATD removal & Mining-based technique & N/A & Not mentioned & 5  \\ \hline
Huang \textit{et al.} \cite{Huang} & 2018 & SATD automatic classification  & Machine learning technique & N/A & design/requirement/defect/test & 8  \\ 
& & & & & documentation & \\ \hline
Liu \textit{et al.} \cite{liu2018satd} & 2018 & SATD detection & Mining-based technique & SATD Detector & design/requirement/defect/test & 8  \\ 
& & & & & documentation & \\ \hline
Lammarino \textit{et al.} \cite{iammarino2019self} & 2019 & SATD removal & Mining-based technique & N/A & Not mentioned & 4  \\ \hline
Farias \textit{et al.} \cite{de2020identifying} & 2020 & SATD identification & Contextualized vocabulary technique & Not mentioned & design/requirement/code/test & 3   \\ 
& & & & & architecture/defect/people/build & \\
& & & & & documentation & \\ \hline

Zampetti \textit{et al.} \cite{zampetti2020automatically} & 2020 & SATD removal & Deep learning technique & SARDELE & Not mentioned & 5  \\ \hline

Xavier \textit{et al.} \cite{xavier2020beyond} & 2020 & SATD identification & Mining-based technique  & Liu \textit{et al.}'s tool \cite{liu2018satd} & design/requirement/code/test & 5  \\ 
& & & & & infrastructure/build/security/UI & \\
& & & & & documentation/performance & \\ \hline
 
\end{tabular}
\end{adjustbox}
\end{table*}

\section{Overview of \approach}
\label{section:approach}

\begin{figure}
    \centering
    \includegraphics[scale=0.6]{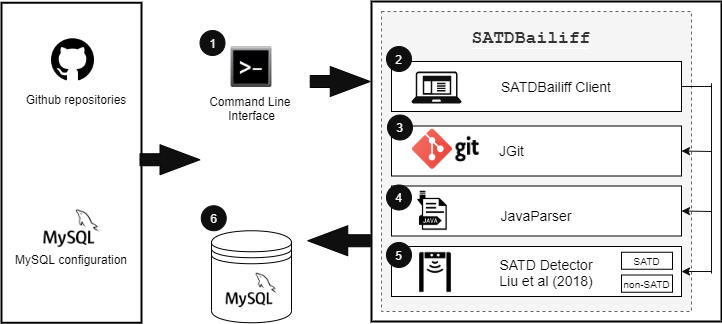}
    \caption{High-level architecture of \approach}
    \label{fig:flow}
\end{figure}

\approach is a Java tool designed to mine the empirical history of SATD instances from Java project Git repositories on a large scale.
This is done with the goal of tracking the additions, removals, and changes to SATD instances that occur during the process of software development. 
\approach's output can be used to better understand the prominence of SATD in software projects at different points over the course of those projects' lifetimes, while also offering new ways to interpret and visualize those SATD instances. While \approach accomplishes its objective using a state-of-the-art classification model \cite{liu2018satd} and a scalable output format, the tool was also designed with modularity in mind to allow for the use of new classification models and output formats.
To collect its data, the tool leverages several existing tools as shown in Figure~\ref{fig:flow}. 

The \approach command line interface (CLI) (Seen in Figure~\ref{fig:flow} as \circled{1}) has two main inputs: A CSV file containing a list of GitHub repositories will be mined by the tool and a MySQL configuration file. The CLI has many other optional inputs to optimally configure the tool.

The Eclipse JGit library\footnote{\url{https://github.com/eclipse/jgit}} \circled{3} is used to collect Java source files from GitHub. This library is also used to collect any commit metadata available, which is output alongside any associated SATD operations found during mining. As explained by Nugroho \textit{et al.} \cite{nugroho2020different}, Git offers a diff utility for users to select various diff algorithms like Myers \cite{Myers} and Histogram \cite{Histogram}, which are utilized to obtain the differences of two identical files located in two different commits. Each algorithm has its own procedures for finding the items presented in the original document but absent in the second one and vice versa. JGit is also used to generate edit scripts between different versions of a project's source code. These edit scripts detail which lines contain removals and additions, and represent all changes to a file between one version of a file to the next. Examples of edit scripts can be seen in Figures~\ref{fig:added_1},~\ref{fig:removed_1},~\ref{fig:false_add_remove}, and~\ref{fig:change_positive}. \approach is configured to use both the Myers \cite{Myers}  and Histogram \cite{Histogram} difference algorithms to generate edit scripts and the tool provides the ability to change between these two algorithms.

JavaParser\footnote{\url{https://github.com/javaparser/javaparser}} \circled{4} is used to extract source code comments from the Java source files obtained by JGit \circled{3}. It also extracts comment metadata, containing method and class, and line numbers. This metadata is output alongside any associated SATD operations found during mining.

The SATD Detector tool presented by Liu \textit{et al.} \cite{liu2018satd} \circled{5} is used for the binary classification of source comments as SATD. This state-of-the-art tool achieved an average F-score of 0.737 during the classification of comments from 5 major open source projects. This classification interface was designed with modularity in mind, and any future higher-performance models can be used with \approach as well.

The logic that bridges all of these tools together is located within the \approach client \circled{2}. The client begins by generating parent-child pairs for every single parent commit found under a given head of the git repository.In this pairing, commits with multiple parents (i.e. merge commits) are not handled, as the current version of the client handles only one branch. Then, for each of those pairs, all source code differences (edit scripts) are calculated for each Java file. All SATD instances are recorded from \textit{each file} impacted by a source code modification in the parent commit, as well as the child commit. A mapping approach is taken to identify which SATD instances may have been impacted by these changes. An SATD instance will map between two commits if both commits contain the same comment, under the same method signature (or lack thereof), and the same containing class name (or lack thereof). SATD instances that share all of those identification properties (e.g., two identical SATD comments in the same method), a number is assigned based on the order they occur. All SATD instances that were not mapped between the two commits are then classified as removed or changed. This classification is determined by the edit scripts generated earlier, and the logic is further described in the Section~\ref{ref:operations}. The result of this process is a complete empirical history of all operations to SATD instances between a given point in a project's lifetime and its origination.

The implicit implementation of \approach outputs to an SQL database \circled{6}, but the tool supports a modular implementation allowing for an extension of other output formats. A simplified data-point sample from the Apache Tomcat project is included in Table~\ref{fig:sample_data}.
The data includes some important features:
\begin{itemize}
    \item \textbf{SATD Id and SATD Instance Id}. Each entry has two identifying integers. The SATD Id is a unique identifier for a single operation to an SATD Instance. An SATD Instance ID is an overarching identifier used to group many SATD operations to a single contiguous instance. In Table~\ref{fig:sample_data}, 
    each entry in this sample would have a different and unique SATD Id.
    \item \textbf{Resolution}. Each SATD operation has a single resolution that impacts the SATD between two commits. These operations include: \textit{SATD\_ADDED}, \textit{SATD\_REMOVED}, \textit{SATD\_CHANGED}, \textit{FILE\_REMOVED}, \textit{FILE\_PATH\_CHANGED}, and \textit{CLASS\_OR\_METHOD\_CHANGED}. The definitions of these operations are described in detail in Section~\ref{ref:operations}.
    \item \textbf{Comment Metadata}. When each SATD operation is recorded, \approach also records the comment's metadata at the time of the operation. This includes data such as the comment type (Line, Block, or JavaDoc as recorded by JavaParser), start and end line, containing class and method, the file name, and the comment itself.
    \item \textbf{Commit Metadata}. When each SATD operation is recorded, \approach also records the metadata of both the child and parent commit. This includes author name and timestamp, committer name and timestamp, and SHA1 commit hash.
\end{itemize}

\begin{table}
    \centering
    \caption{Simplified sample data from the Apache Tomcat project}
    \begin{tabular}{|p{4.5em}|p{9em}|p{12em}|p{3.6em}|p{10em}|}
     \hline
      \textbf{SATD\_id} & \textbf{SATD\_instance\_id} & \textbf{resolution} & \textbf{commit} & \textbf{comment}\\
     \hline
    13958 &  652915385 & SATD\_ADDED   & 09b640e & TODO: 404\\
     14317 &  652916048 & FILE\_PATH\_CHANGED & decfe2a & TODO: 404\\ 
     13665 &  652915615 & FILE\_REMOVED & a457153 & None\\
    
     \hline
    \end{tabular}
    
    \label{fig:sample_data}
\end{table}

\noindent{\textbf{Operations on SATD}}
\label{ref:operations}

Previously, Maldonado \textit{et al.} \cite{Maldonado2017-2} conducted an empirical study on the removal of SATD that pointed changes to SATD instance incorrectly as removals and additions. 
In addition to mistaking file renames, this would detect instances like the examples in Figures~\ref{fig:false_add_remove} and ~\ref{fig:change_positive} as having both resolved the original SATD instance and added the new version to the project, respectively.

\approach resolves this issue by handling SATD comment and file name changes as operations in-between additions and removals of SATD. In order to observe a more fine-grained change in source code changes, the tool observes edit scripts for changes to specific lines of code made between each commit. Edit scripts detail the addition and removal of specific lines of code within a file. An example edit script can be seen in Figure~\ref{fig:false_add_remove} detailed by the red and green highlighted source text.

\begin{figure}
    \centering
    \includegraphics[scale=0.4]{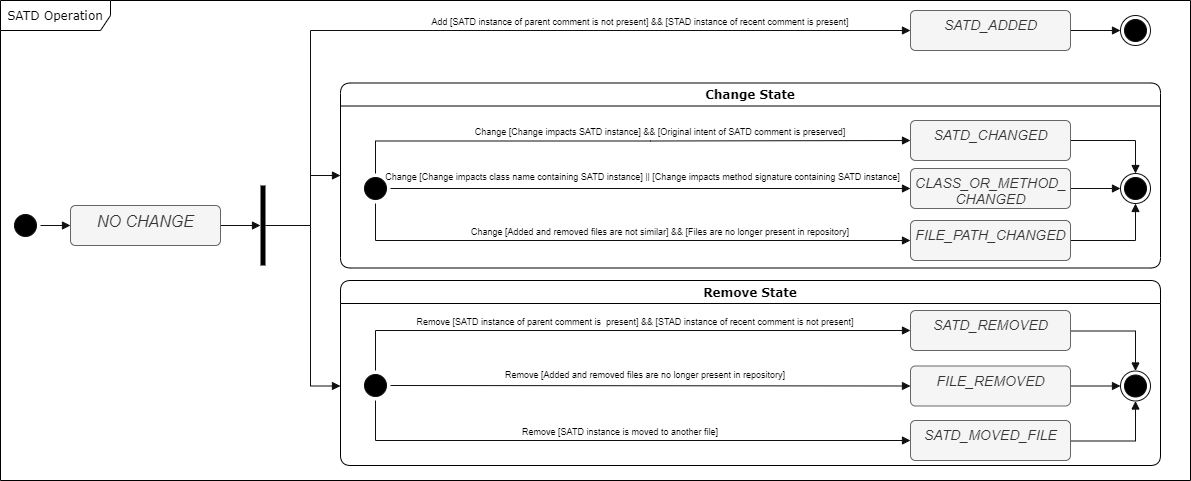}
    \caption{SATD operations captured by \approach}
    \label{fig:statechart}
\end{figure}

\begin{figure}[h]
\centering 
\begin{tikzpicture}
\begin{scope}[scale=0.85]
\pie[rotate=180,text=legend]{42/SATD\_ADDED,24.5/SATD\_REMOVED,14.2/FILE\_PATH\_CHANGED,7.4/FILE\_REMOVED,6.8/CLASS\_OR\_METHOD\_CHANGED,3.5/SATD\_CHANGED,1.6/SATD\_MOVED\_FILE}
\end{scope}
\end{tikzpicture}
\caption{Distribution of SATD operations in the studied 5 projects}
\label{fig:satdoperation}
\end{figure}
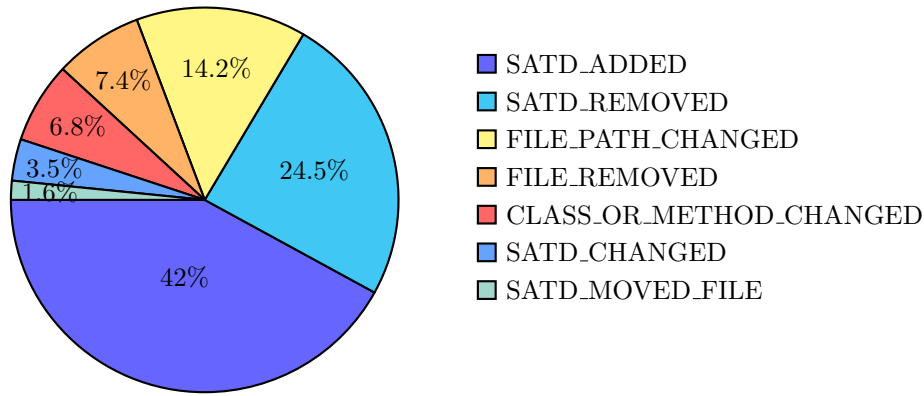

Figure~\ref{fig:statechart} depicts the SATD operations captured by our tool and Figure~\ref{fig:satdoperation} shows the distribution  of these operations in our five projects. Our tool captures the following cases of the SATD removal: (1) when the file has been deleted-- that is, when SATD disappears because the related code is no longer in the system (i.e., accidental removal), and (2) when SATD comments are removed but the code still exists. 
It is worth noting that the purpose of the tool is not to evaluate the effectiveness of SATD removal or to explore whether SATD is addressed or not. Our study supports the existing studies that focus on these aspects by providing more details and explanations related to the evolution of SATD (e.g., method changed or not, file removed or not, track when SATD starts and when it is removed, etc). 
The cases that we plan to capture in the extended version of the tool are: (1) when the SATD comment is removed entirely, and a new SATD comment is added instead, although we have not found any such case based on our manual analysis, (2) when SATD comment has dropped while the code still remains unchanged, and (3) when SATD comments are removed and there are changes in the method.

The next subsections describe the process of identifying each of the operations that \approach handles. The sections use the variables:

\begin{itemize}
    \item $C_a, C_b$, the \textbf{parent commit} and the more \textbf{recent commit}, respectively.
    \item $S_a$, $S_b$, a specific \textbf{SATD instance} in the parent commit and the more recent commit, respectively. It should be assumed that $S_a$ and $S_b$ are intentionally related to each other if not identical.
    \item $SA_a, SA_b$, an \textbf{arbitrary other SATD instance} in the same file unrelated to $S_a$ or $S_b$ in commits $C_a$ and $C_b$ respectively.
    \item $E_{1}, E_{2}, ... E_{n}$, the \textbf{edit scripts} generated when differencing $C_a$ and $C_b$ that impact the lines of SATD Comment $S_1$. Multi-line SATD comments may have multiple edit scripts that impact it where $n$ is used to differentiate these line-based edit scripts.
\end{itemize}

\begin{enumerate}
    \item \textbf{SATD\_ADDED \& SATD\_REMOVED}


A naive comment differencing algorithm determines \textbf{SATD\_ADDED} instances would exist in any $C_a$ where $S_a$ is not present and the associated $C_b$ where $S_b$ is present. This satisfies a basic case in Figure~\ref{fig:added_1}. 

\begin{figure}
    \centering
    \begin{tabular}{|p{25.2em}|}
    \hline
        \begin{lstlisting}
    body = exchange.getOut().getBody();
|\hladd{+~~// TODO: what if exchange.isFailed()?~~}|
    if (body != null) {
        \end{lstlisting}
    \\\hline
    \end{tabular}
    \caption{A basic case \textbf{SATD\_ADDED} instance}
    \label{fig:added_1}
\end{figure}

However, SATDBaillif needs to account for \textit{changes} in SATD comments. In Figure~\ref{fig:false_add_remove}, these changes would be identified by separate \textbf{SATD\_REMOVED} and \textbf{SATD\_ADDED} instances using the naive logic. Instead, SATDBAllif determines that if a single edit script $E_n$ exists such that $E_n$ impacts $S_b$ without impacting $SA_a$, then $S_b$ was added by $C_b$.

A naive comment differencing algorithm also determines \textbf{SATD\_REMOVED} instances would exist in any $C_b$ where $S_b$ is not present and the associate $C_a$ where $S_a$ is present. This satisfies the basic case in Figure~\ref{fig:removed_1}.

\begin{figure}
    \centering
    \begin{tabular}{|p{25.2em}|}
    \hline
        \begin{lstlisting}
  protected void connectIfNecessary() {
|\hlremove{-~~ // can we avoid copy-pasting?~~~~~~~~~}|
    if (!client.isConnected()) {
        \end{lstlisting}
    \\\hline
    \end{tabular}
    \caption{A basic case \textbf{SATD\_REMOVED} instance}
    \label{fig:removed_1}
\end{figure}

However, in the case of Figure~\ref{fig:false_add_remove}, it is seen that a more robust algorithm must be used to detect SATD removals. \approach handles this case such that if a single edit script $E_n$ exists such that $E_n$ impacts $S_a$ without impacting $SA_b$, then $S_a$ was removed by $C_b$. 

It can also be the case that the previous logic used by \approach to classify additions and removals to be false, and for the tool to still classify an operation as an \textbf{SATD\_ADDED} or \textbf{SATD\_REMOVED}. This case is better identified by the logic in the following \textbf{SATD\_CHANGED} section. 

\begin{figure}
    \centering
    \begin{tabular}{|p{25.3em}|}
    \hline
        \begin{lstlisting}
    logger.log(\say{Init successful});
|\hlremove{-~~ // Moved this config to the bottom~~~~}|
|\hladd   {+~~ // Moved this config~~~~~~~~~~~~~~~~~~}|
|\hladd   {+~~ // to the bottom~~~~~~~~~~~~~~~~~~~~~~}|
    super.init();
        \end{lstlisting}
    \\\hline
    \end{tabular}
    \caption{A case of a would-be false \textbf{SATD\_ADDED} and \textbf{SATD\_REMOVED} instances}
    \label{fig:false_add_remove}
\end{figure}

\item \textbf{SATD\_CHANGED}
\label{subsubsecchanged}

Changes in an SATD comment can be difficult to determine because they can remove the SATD comment entirely or replace it with a new non-SATD, or irrelevant, comment. Ideally, only SATD comments which preserve the original intent of the SATD comment should be recorded as a \textbf{SATD\_CHANGED}. However, this requires the comprehension of the comment and its relationship with the TD described. Any updates on the comment text may be related to the addition or deletion of more details, as seen later in the manual validation. Since we cannot assess whether an update to an SATD means partially or totally addressing it, our tool will report any form of textual change to the SATD, as \textbf{SATD\_CHANGED}. To do so, the edit scripts of the file are first observed. If a single edit script $E_n$ exists such that $E_n$ impacts both $S_a$ and $SA_b$, then it can be determined that a change may have occurred. Our logic currently flags as \textbf{SATD\_CHANGED} additions of newlines (see Figure~\ref{fig:false_add_remove}), spelling corrections (Figure~\ref{fig:change_positive}), addition or removal of adjacent related and unrelated comments, and URL updates. 
\begin{sloppypar}
\approach also checks the updated comments to determine if they still can be classified as SATD instances. Figure~\ref{fig:change_removal} shows an example of an SATD instance which is recorded in $C_a$ as \lstinline{"lets test the receive worked\nTODO"} due to how the tool groups adjacent comments. In $C_b$, the removal of the \lstinline{\nTODO} substring of the instance results in the instance no longer being classified as SATD, and thus \approach reports this instance as \textbf{SATD\_REMOVED}. 
Without this additional verification, this SATD instance would be incorrectly reported as having only been changed. However, this approach has its own limitation: if a developer removes one SATD and adds an another entirely unrelated SATD, for the same method, in the same commit, like in Figure~\ref{fig:change_negative}, our tool would flag this as \textbf{SATD\_CHANGED}. Fortunately, this case does not seem to be frequent, as shown later in the manual validation, the precision of our model is up to \textbf{96\%}. Yet, we report the limitation of our tool in Section\ref{section:limitations}, and we discuss our current explorations to address it.
\end{sloppypar} 



\begin{figure}
    \centering
    \begin{tabular}{|p{25.2em}|}
    \hline
        \begin{lstlisting}
  try {
|\hlremove{- ~~// Maybe this already exi}\hlremovechange{tst}\hlremove{~~~~~~~~~~}|
|\hladd   {+ ~~// Maybe this already exi}\hladdchange{sts}\hladd{~~~~~~~~~~}|
    success = client.changeDir(dirName);
        \end{lstlisting}
    \\\hline
    \end{tabular}
    \caption{A valid \textbf{SATD\_CHANGED} instance}
    \label{fig:change_positive}
\end{figure}

\begin{figure}
    \centering
    \begin{tabular}{|p{25.2em}|}
    \hline
        \begin{lstlisting}
  c = endpoint.createChannel(session);
|\hlremove{- // TODO: what if creation fails?~~~~~~~~}|
|\hladd   {+ // Bug 1402~~~~~~~~~~~~~~~~~~~~~~~~~~~~~}|
  c.connect();
        \end{lstlisting}
    \\\hline
    \end{tabular}
    \caption{A possibly valid but false \textbf{SATD\_CHANGED} instance}
    \label{fig:change_negative}
\end{figure}


\begin{figure}
    \centering
    \begin{tabular}{|p{25.2em}|}
    \hline
        \begin{lstlisting}
  // lets test the receive worked
|\hlremove{- // TODO~~~~~~~~~~~~~~~~~~~~~~~~~~~~~~~~~}|
|\hlremove{- // assertMessageRec("???@localhost");~~~}|
|\hladd   {+ assertMessageRec("copy@localhost");~~~~~}|
  c.connect();
        \end{lstlisting}
    \\\hline
    \end{tabular}
    \caption{\textbf{SATD\_REMOVED} instance removing only part of a comment}
    \label{fig:change_removal}
\end{figure}

\item \textbf{CLASS\_OR\_METHOD\_CHANGED}

The final edit script source code change detected by \approach is modification to an SATD instance's containing class or method. These cases are detected if any $E_n$ impacts the class or method containing $S_a$ such that the method signature or the class name are changed. \approach does not currently identify when SATD is moved throughout a file by multiple separate edit scripts.

\item \textbf{FILE\_REMOVED \& FILE\_PATH\_CHANGED}

File removals are detected implicitly by Git, where a similarity between added and removed files determines whether a file is removed or renamed when it is no longer present in the repository when committed. This detection method was available as part of the JGit library, and was utilized for identifying \textbf{FILE\_REMOVED} and \textbf{FILE\_PATH\_CHANGED} instances.

\end{enumerate}
\section{\approach Usage}
\label{section:usage}

This section describes the usage of \approach and its features.
\subsection{Installation}
The most up-to-date precompiled binaries can be found on the project's GitHub\footnote{\url{https://github.com/smilevo/SATDBailiff}} repository and on the tool's website.\footnote{\url{https://smilevo.github.io/self-affirmed-refactoring/SCP20_index.html}} The project can be run using a Java version 8+ and is otherwise OS independent as it could be run through Docker. 
\subsection{Usage}
\approach can be used either through its Command Line Interface (CLI) or Application Program Interface (API), and can be easily modified to support different output types and classification models.

The \approach Command Line Interface (CLI) (Seen in Figure~\ref{fig:flow} as \circled{1}) has two main inputs: A CSV file containing a list of GitHub repositories and a MySQL configuration file. The CLI has many other optional inputs to optimally configure the tool.

The CSV file containing repository information details which repositories will be mined by the tool, and where the mining will terminate. A terminal commit value can be added next to the repository URI to add an terminal point in time to which the tool can mine. This is done primarily to assure reproducibility between datasets mined at different times. If absent, the terminal commit value will default to the most recently available commit in the repository. The output format of the data also allows for a manual filtering of SATD operations by date. However, it should be noted that un-merged branches of the repository at certain timestamps are likely to cause difference between a pre- and post-execution commit filtering. Git credentials for private repositories can be added as a separate program argument.

By default, \approach outputs to a MySQL database. The tool intends for a MySQL database to be set up to receive the system's output. Configuration fields for this database must be supplied to the program to connect to the database. A description for the required fields, as well as the required schema for the database can be found in the GitHub repository.

Other runtime variables available within \approach include: 
\begin{itemize}
    \item File differencing algorithms available through JGit - currently only Histogram and Myers;
    \item The Normalized Levenshtein distance threshold (between 0 and 1) described in Section~\ref{section:approach};
    \item A toggle for error display;
    \item A help menu display;
\end{itemize}

When run, \approach will display using the interface in Figure~\ref{fig:runtime_snapshot}. This output includes a detailed description of the runtime duration of the tool, the number of commits differences, and a description of each entry that caused any sort of detectable error in the system, if errors display is toggled. If error display is disabled, then only the number of errors encountered will be displayed.

\begin{figure*}
    \centering
        \begin{small}
            \begin{verbatim}
$ java -jar SATDBailiff.jar -d mySQL.properties -r repos.csv
Completed analyzing 78 diffs in 26,103ms (334.65ms/diff, 0 errors) - analogweb/core
3wks/thundr -- Mining SATD  (13.3%, 46/346, 0 errors) - e048736
            \end{verbatim}
        \end{small}
        
    \caption{Sample runtime snapshot}
    \label{fig:runtime_snapshot}
\end{figure*}



\subsection{Interpreting Output}

\begin{figure}
    \centering
    \includegraphics[width=0.6\linewidth]{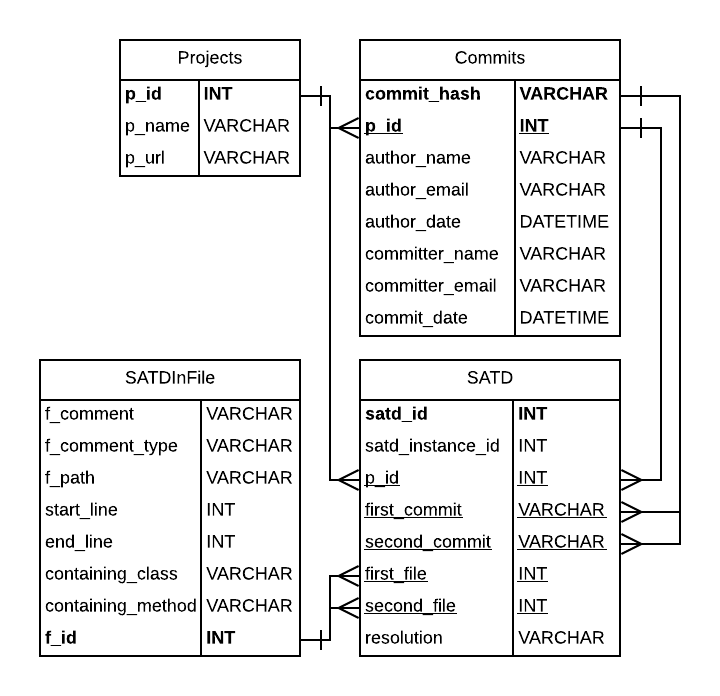}
    \caption{\approach output schema}
    \label{fig:schema}
\end{figure}

The released implementation of \approach outputs to a SQL database. The schema for the output is detailed in Figure ~\ref{fig:schema}. Project table contains the list of open-source Java projects hosted on GitHub that are utilized in the study. Commits table stores commit level metadata by looping through the commit log for each studied project. The metadata includes author name and timestamp, committer name
and timestamp, and SHA1 commit hash. A high level information about SATD is stored in the SATD table, whereas SATD comment’s metadata including the comment type, start and end line, containing class and method, the file name, and the comment itself is recorded in SATDInFile table.

\approach  also generates a CSV file and an HTML file that contain the same information saved in the database. The CSV file is helpful for results parsing, while the HTML helps with the visualization of the results.

\section{\approach Applicability}
\label{section:applicability}

Tracking SATDs are vital for maintaining healthy software systems and reducing maintenance cost. As reported by Wehaibi \textit{et al.} \cite{wehaibi2016examining}, SATD has negative implications on the software development process since it makes it hard to perform any future changes. Our tool could serve as a guide; helping developers pay technical debt by recommending the changes they should perform. Tracking the additions, removals, and changes to SATD instances that occur during the process of software development can be used to better understand the prominence of SATD in software projects at different points over the course of those projects’ lifetimes. Additionally, tracking SATD instances would provide valuable support for existing studies by capturing the removals without the need to manually validate them and will help future research to explore how SATD is addressed and how SATD removal should be measured (e.g., evaluate the effectiveness of the SATD removal when a method has changed along with the removal). Additionally, our tool can assist in exploring ways to reduce/manage technical debt. For instance, a recent study \cite{iammarino2019self} shows that  there is a higher chance for refactoring actions to occur together with SATD removals than with other changes. Another recent study on refactoring documentation \cite{alomar2019can,alomar2020toward,alomar2021we} shows that developers reported \say{Fix Technical Debt} in their commit messages which shows that refactoring actions could be used to cope with technical debt.

Martini et al. \cite{martini2018technical} conducted an industrial case study with practitioners and their results show that 25.9 \% of the average development time and effort is spent by 215 practitioners to manage technical debt. In their survey, they found that  some respondents report spending more than 40\% of their time managing technical debt. However, only 26\% of the participants used a tool to track technical debt. The respondents mentioned the tools used to track technical debt are: comments, documentation, issues, backlog, static analyzer, lint and test coverage. Backlogs (i.e.,  Jira, Hansoft, and Excel) are the most used tool among the participants. According to their survey with the practitioners, respondents show awareness of indicators of technical debt like comments or documentation. This sheds light on the need of developing tools to help with automatically tracking technical debt early in the development stages. Here are some examples that showcase the usefulness of the tool:

\begin{itemize}
    \item \textbf{Promoting the adoption of tracking SATD in practice}. Fixing bugs or adding new features would be challenging and a time consuming task  as the system evolves, and software engineers might experience a negative impact on the quality. Adopting technical debt tracking processes by managers or experienced developers, who understand the importance of technical debt, reduces maintenance cost. Our tool is one of the methods that helps promote the adoption of tracking SATD practice. This feature would be fully achieved with future refinement of the tool, once we gather more feedback from researchers and practitioners who are using the tool.
    \item \textbf{Indicating refactoring technique that removed the SATD}. Since the management of technical debt has been heavily correlated with refactoring, the tool shows the history of SATDs and the refactoring operations that helped remove the debt. This assists in analyzing the necessary changes needed to remove technical debt.
\end{itemize}

\section{\approach Validation}
\label{section:validation}

 \subsection{Internal Validation:}
To verify the accuracy of \approach, a manual analysis was performed on a sample of 1,882 entries mined from 5 large open source Java projects described in Table \ref{tab:Studied Projects}, each as their own strata. The number of samples taken from each project is determined by the total number of SATD instances mined from the projects. Each of the SATD instances selected includes SATD operations (removed or changed) performed on a single instance of SATD. Each instance in the sample will represent an entirely unique instance of SATD. A simplified example of a single SATD instance can be seen in Table~\ref{fig:sample_data}. The results of this analysis can be seen in Tables~\ref{fig:manual_analysis_removed} and~\ref{fig:manual_analysis_changed}.


For the tool validation, we selected five projects that are widely used by the state-of-the-art research in the context of SATD \cite{Maldonado2017-2,Zampetti,iammarino2019self,zampetti2020automatically}, as these projects are claimed to have an adequate number of SATD instances. We build on top of findings of Maldonado \textit{et al.} \cite{Maldonado2017-2} and start from their SATD removal dataset. In addition to the publication of this tool, an empirical history of SATD instances in 30 open source software projects, produced by \approach, are available on our website.

The detail of the studied 30 projects is summarized in Table \ref{tab:Studied Projects_30projects}. Figure \ref{fig:satdoperation-30projects} shows the distribution of SATD operations captured by our tool in the 30 projects. We observe that the dominant category is the SATD\_REMOVED with 77,29\%. Our future work involves exploring the technique developers used to remove SATDs from their projects.

\approach was configured to only mine SATD instances between the original commit to each project, and the most recent commit reported by the Maldonado \textit{et al.} study.

\begin{table}[h]
    \centering
    \caption{Details of the studied 30 projects.}
    \begin{tabular}{|l|r|r|r|}
     \hline
     \textbf{Project} &  \textbf{\# Commits} &  \textbf{\# Contributors} & \textbf{\# detected SATD} \\
     \hline 
gitools/gitools \footnote{\url{	https://github.com/gitools/gitools}} &  2,051 & 4 &  1,704  \\
twilio/twilio-java	 \footnote{\url{https://github.com/twilio/twilio-java}} & 118 & 3 &  89  \\
hbutani/sqlwindowing	 \footnote{\url{https://github.com/hbutani/sqlwindowing}} & 424 & 2&  26  \\
pulse00/twig-eclipse-plugin	 \footnote{\url{https://github.com/pulse00/twig-eclipse-plugin}}&  499 & 4 & 359 \\
eclipse/vert.x	 \footnote{\url{https://github.com/eclipse/vert.x}} & 4,950 & 196 & 452  \\
ngdata/lilyproject	 \footnote{\url{https://github.com/ngdata/lilyproject}} & 2,717 & 7 & 3,908\\
hibernate/hibernate-validator	 \footnote{\url{https://github.com/hibernate/hibernate-validator}} & 4,198 & 78 & 885 \\
wocommunity/wonder	 \footnote{\url{https://github.com/wocommunity/wonder}} & 13,885 & 70 & 2,215 \\
apache/maven-plugins	 \footnote{\url{https://github.com/apache/maven-plugins}} & 15,063 & 36 & 2,300 \\
davemckain/qtiworks	 \footnote{\url{https://github.com/davemckain/qtiworks}} & 2,065 & 2 &  635 \\
icy-imaging/icy-kernel	 \footnote{\url{https://github.com/icy-imaging/icy-kernel}} & 665 & 5 & 978  \\
jfxtras/jfxtras-labs	 \footnote{\url{https://github.com/jfxtras/jfxtras-labs}} & 1,834 & 26 & 3,136   \\
romanchyla/montysolr	 \footnote{\url{https://github.com/romanchyla/montysolr}} & 1,628 & 6 & 927   \\
apache/httpclient	 \footnote{\url{https://github.com/apache/httpclient}} & 3,225 & 51 &  690  \\
socialsoftware/blended-workflow	 \footnote{\url{https://github.com/socialsoftware/blended-workflow}} & 849 & 6&  824 \\
crosswire/jsword	 \footnote{\url{https://github.com/crosswire/jsword}} & 1,865 & 8 & 1,275 \\
motech/motech-whp	 \footnote{\url{https://github.com/motech/motech-whp}} & 2,190 & 14 &  97  \\
eclipse/bpmn2-modeler	 \footnote{\url{https://github.com/eclipse/bpmn2-modeler}} & 1,442 & 6 & 1,072\\
eclipse/ecf	 \footnote{\url{https://github.com/eclipse/ecf}} & 11,793 & 11 & 1,559   \\
getrailo/railo	 \footnote{\url{https://github.com/getrailo/railo}} & 3,990 & 15 & 862 \\
msbarry/xtest	 \footnote{\url{https://github.com/msbarry/xtest}} & 341 & 1 & 367   \\
projectdanube/xdi2	 \footnote{\url{https://github.com/projectdanube/xdi2}} & 2,504 & 3 & 254 \\
scribble/scribble-java	 \footnote{\url{https://github.com/scribble/scribble-java}} & 2,139 & 4 & 6,803 \\
tinkerpop/gremlin	 \footnote{\url{https://github.com/tinkerpop/gremlin}} & 1,227 & 15 &  159 \\
adangel/pmd	 \footnote{\url{https://github.com/adangel/pmd}} & 17,691 & 99 & 3,997  \\
qcadoo/mes	 \footnote{\url{https://github.com/qcadoo/mes}} & 14,906 & 30 & 1,440  \\
belaban/jgroups	 \footnote{\url{https://github.com/belaban/jgroups}} & 19,213 & 72 & 1,193  \\
merks/xcore \footnote{\url{https://github.com/merks/xcore}} & 118 & 3 &  89,175   \\
rvonmassow/xdoc	 \footnote{\url{https://github.com/rvonmassow/xdoc}} & 151 & 548 & 6
   \\
nasa/certware	 \footnote{\url{https://github.com/nasa/certware}} & 237 & 4 & 1,091   \\

     \hline
    \end{tabular}
    
    \label{tab:Studied Projects_30projects}
\end{table}

\begin{figure*}[h]
\centering 
\begin{tikzpicture}
\begin{scope}[scale=0.85]
\pie[rotate=180,text=legend, outside under=10]{10.12/SATD\_ADDED,77.29/SATD\_REMOVED,5.07/FILE\_PATH\_CHANGED,3.03/FILE\_REMOVED,2.05/CLASS\_OR\_METHOD\_CHANGED,2.06/SATD\_CHANGED,0.40/SATD\_MOVED\_FILE}
\end{scope}
\end{tikzpicture}
\caption{Distribution of SATD operations in the studied 30 projects.}
\label{fig:satdoperation-30projects}
\end{figure*}

\begin{table}[h]
    \centering
    \caption{\approach manual analysis results for SATD removal}
    \begin{tabular}{ |p{4em}|R{5em}|R{5.2em}|R{4.8em}|   }
     \hline
     \textbf{Project} & \textbf{\# Entries} & \textbf{\# False Positive} & \textbf{Precision}\\
     \hline 
    Camel          & 20  & 1 & 0.99\\
     Gerrit         & 284  & 6  & 0.99\\
     Hadoop         & 608  & 3  & 0.99\\
     Log4j          & 7  & 0   & 1.00\\ 
     Tomcat         & 432  & 8  & 0.99\\
     \hline
     \textbf{Total} & 1351 & 18 & \textbf{0.99}\\ 
    
     \hline
    \end{tabular}
    
    \label{fig:manual_analysis_removed}
\end{table}

\begin{table}[h]
    \centering
    \caption{\approach manual analysis results for SATD changed}
    \begin{tabular}{ |p{4em}|R{5em}|R{5.2em}|R{4.8em}|   }
     \hline
     \textbf{Project} & \textbf{\# Entries} & \textbf{\# False Positive} & \textbf{Precision}\\
     \hline 
     Camel          & 19  & 2 & 0.99\\
     Gerrit         & 223  & 25 & 0.92\\
     Hadoop         & 112  & 13  & 0.96\\
     Log4j          & 9  & 0   & 1.00\\ 
     Tomcat         & 168  & 24 & 0.93\\
     \hline
     \textbf{Total} & 531 & 64 & \textbf{0.96}\\ 
    
     \hline
    \end{tabular}
    
    \label{fig:manual_analysis_changed}
\end{table}

\begin{figure}[htbp]
    \centering
    \begin{tabular}{|p{28em}|}
    \hline
        \begin{lstlisting}
protected void processSoapConsumerOut
(Exchange exchange) throws Exception {
LOG.info("processSoapConsumerOut:" + exchange);
|\hlremove{- // TODO~~~~~~~~}|
|\hladd   {+ // TODO check if the message is oneway message
             // Get the method name form the soap endpoint}|
  ...
 }
        \end{lstlisting}
    \\\hline
    \end{tabular}
    \caption{An expansion case of \textbf{SATD\_CHANGED} instance}
    \label{fig:change_expansion}
    
\vspace{0.70cm}

    \centering
    \begin{tabular}{|p{28em}|}
    \hline
        \begin{lstlisting}
 public void onExchange(HttpExchange exchange)
 {
|\hlremove{-   // we need an external HTTP client such as commons-httpclient
               // TODO}|
|\hladd   {+ // TODO}|
 }
        \end{lstlisting}
    \\\hline
    \end{tabular}
    \caption{An abbreviation case of \textbf{SATD\_CHANGED} instance}
    \label{fig:change_abbreviation}

\vspace{0.70cm}
    \centering
    \begin{tabular}{|p{28em}|}
    \hline
        \begin{lstlisting}
|\hlremove{-  //TODO: YARN-3284
              //The containerLocality metrics will be exposed from AttemptReport}|
|\hladd   {+ // TODO:YARN-3284}|
  private void createContainerLocalityTable
  (Block html) {
  ...
  }
        \end{lstlisting}
    \\\hline
    \end{tabular}
    \caption{A generalization case of \textbf{SATD\_CHANGED} instance}
    \label{fig:change_generalization}    
    
\vspace{0.70cm}
    \centering
    \begin{tabular}{|p{28em}|}
    \hline
        \begin{lstlisting}
|\hlremove{- // TODO fix this test}|
|\hladd   {+ // TODO fix this test, it looks like AMQP don't support Object message}|
public void xtestJmsRouteWithObjectMessage() 
throws Exception {
        ...
 }
        \end{lstlisting}
    \\\hline
    \end{tabular}
    \caption{A specialization case of \textbf{SATD\_CHANGED} instance}
    \label{fig:change_specialization}  
\end{figure}

\begin{figure}
    \centering
    \begin{tabular}{|p{28em}|}
    \hline
        \begin{lstlisting}
public void afterPropertiesSet() {
|\hlremove{- // TODO: is needed when we add support for when predicate}|
if (getOutputs().size() == 0) {
|\hladd   {+ // no outputs}|
return;
}

        \end{lstlisting}
    \\\hline
    \end{tabular}
    \caption{A case of \textbf{SATD\_REMOVED} instance when there is no overlap}
    \label{fig:remove_entirelydifferent}  
\end{figure}

To perform the analysis, one of the authors was given the set of SATD instances and asked to locate the exact location of each of the SATD operations using the GitHub website. A \say{correct} entry was identified as an entry in which every operation made to the SATD instance could be located using the GitHub interface. Any unnecessary additional, missing, or inaccurate operations found on GitHub would result in the entire entry being incorrect. For entries that were not removed from the project, their existence in the terminal commit supplied to \approach  was confirmed. For transparency of this analysis, a GitHub link to the exact source modification was recorded in each of the projects where available. These results are available on the project's website.\footnote{\url{https://smilevo.github.io/self-affirmed-refactoring/SCP20_index.html}} During validation, it was assumed that all binary classifications of source code comments as SATD were correct.

The results of the manual analysis (Tables~\ref{fig:manual_analysis_removed} and~\ref{fig:manual_analysis_changed}) find \approach  to have a precision of 0.99 and 0.96 for SATD removal and SATD changes, respectively. 
It is infeasible to always achieve a perfect precision of the results extracted from the tool due to the inconsistent nature of open source projects and their development practices. Open source projects are varied in size, contributors,  number of comments, and SATDs. Thus, the performance of the tool might differ based on development practices of the selected projects. While a higher level of accuracy could have been achieved, it should be noted that many of the incorrect instances were partially correct. For example, instances frequently were found to be incorrect because they became dissociated with one another, where a connection between an SATD instance's addition to the project and its deletion from the project was not made by the tool. In cases where only the additions or removals are observed from the dataset, the accuracy of the data provided by the tool is much more reliable.

Difficulties in solving many of the tool's issues came from the imperfect nature of working with edit scripts produced by Git differencing tools. Edit scripts are used to show an algorithm's best guess of changes in files inside of a Git repository, and do not always reflect the true intentions of the developer who made them \cite{Frick}. An example of an edit script can be seen in Figure~\ref{fig:change_positive} depicted as the red and green highlight used to represent a source code change. JGit uses the Myers \cite{Myers} and the Histogram \cite{Histogram} diff algorithm to produce edit scripts. \approach provides the ability to change between these two algorithms, and the Myers algorithm was used during performed manual validation. Both of these algorithms maintain a manually validated accuracy of less than 0.9 \cite{Frick}. While, in many cases, an invalid edit script will not directly invalidate \approach's ability to identify operations to SATD instances, this inaccuracy still serves as a significant limitation in the upper bound of accuracy achievable by this tool.

To assure that these errors are not able to silently pollute the dataset, the tool reports any known errors that are encountered during the mining process.
This workaround was taken as an optimistic precaution for an issue that may not have a perfect alternative solution.
For example, SATD that is added during a merge commit which was not present in either of the merge branches is not detected with an SATD\_ADDED entry. If that SATD is modified or removed later, then the entry would be added to the project before the SATD\_ADDED entry was found. Because the search occurs chronologically starting with the oldest commit in the project, the system can detect this as an issue and will output an error to the terminal during runtime.

When performing the manual validation, we noticed that developers changed SATD comments  as follows:
\begin{itemize}
    \item An expansion form of SATD comment: a comment that has more explanation of the SATD case. For instance, in Figure \ref{fig:change_expansion}, the comment contains more than just the TODO tag, as the developer added a couple of functionalities to implement, as part of the TODO.
    \item An abbreviation form of SATD comment: the opposite of the expansion in which a shortened form of SATD comment is provided. For instance, in Figure \ref{fig:change_abbreviation}, the text explaining the task, tagged with the TODO, has been removed, however, the TODO has been kept, and therefore, the comment is still flagged as SATD. 
    \item A generalization form of SATD comment: a comment has broad SATD context.  In Figure \ref{fig:change_generalization}, the details of the TODO have been removed, and the reference to the active issue has been kept.
    \item A specialization form of SATD comment: the opposite of the generalization in which a comment has focused SATD context. For example, in Figure \ref{fig:change_specialization}, details on how to fix the test and make it support the Object message, have been added. 
\end{itemize}
As for the SATD removal case, we noticed the case in which there is entirely no overlap between the SATD comments as shown in Figure \ref{fig:remove_entirelydifferent}. 

 \subsection{External Validation:}
 
To further assess the usefulness of the tool, we perform an external validation by involving 15 graduate students from the Department of Software Engineering and Data Science at Rochester Institute of Technology. All participants volunteered to participate in the experiment. For each participant, the experiment consists of (1) randomly choosing an Java-based open-source project and fork it, (2) detecting 10 issues in the open-source project, using PMD and/or SpotBugs, (3) writing a comment, describing the issue detected inside the method, while making sure the comment is written in the form of SATD, then committing it (4) addressing the issues one by one by making the necessary code changes recommended by PMD and/or SpotBugs, while making sure to remove the comment describing it, then committing the changes (5) running SATDBailiff tool to identify these added and removed comments, and finally (6) filling out the survey to give feedback about the tool. The idea behind this experiment is to identify potential TD instances, and document their existence, then addressing them and removing their corresponding comments. We used PMD and SpotBugs because they are known to be good indicators for technical debt in the source code \cite{zampetti2017recommending,amanatidis2020evaluating}. Upon the completion of the experiment, we calculated the average recall, and we found the average recall is 0.90, which is considered acceptable. The replication package for the whole experiment is available in \footnote{\url{https://smilevo.github.io/self-affirmed-refactoring/SCP20_index.html}}.
 
The survey consisted of 6 questions that are divided into 2 parts. The first part of the survey includes demographics questions about the participants. In the second part, we asked about the (1)  satisfaction of the tool aspects, (2) preferred features of the tool, (3) suggested features of the tool to be added in the future, and (4) any comments about the tool. We constructed the survey to use 5-point ordered response scale (“Likert scale”) question on the aspects of the tool, and 3 open-ended questions on the preferred features, suggested features, and general comments about the tool.
 
\begin{figure}[h]
\centering 
\begin{tikzpicture}
\begin{scope}[scale=0.85]
\pie[rotate=180,text=legend]{26.7/less than 1 year,53.3/1 - 3 years,13.3/3 - 5 years,6.7/more than 5 years}
\end{scope}
\end{tikzpicture}
\caption{Participant programming experience
in years.}
\label{fig:experience}
\end{figure}
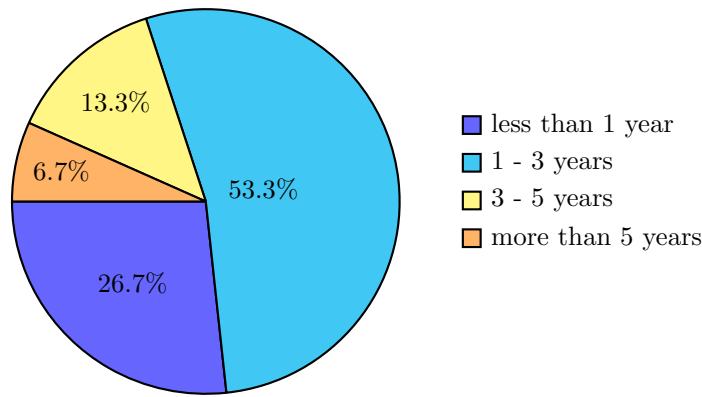

  As shown in Figure \ref{fig:experience}, the experience of these participants with programming ranged from less than a year to more than 5 years. As for the familiarity with the concept of technical debt, 66.7\% of the participants are familiar with the technical debt, whereas 33.3\% are not familiar with the concept of technical debt. Prior to the execution process, the participants were provided with a 75-minute tutorial on technical debt along with reference materials.
  
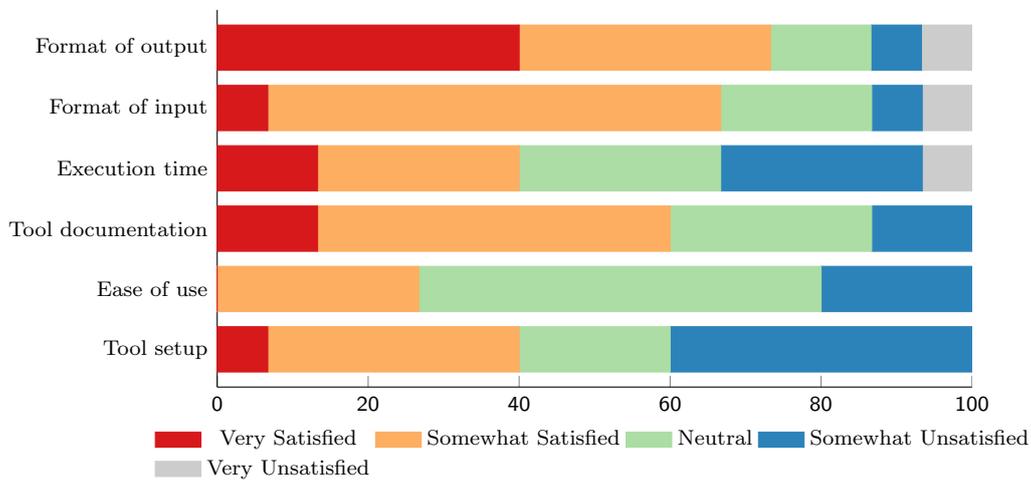
\begin{figure*}[h]
\centering
\begin{tikzpicture}
\begin{axis}[
    xbar stacked,
    legend style={
    legend columns=4,
        at={(xticklabel cs:0.5)},
        anchor=north,
        draw=none
    },
    ytick=data,
    axis y line*=none,
    axis x line*=bottom,
    tick label style={font=\footnotesize},
    legend style={font=\footnotesize},
    label style={font=\footnotesize},
    xtick={0,20,...,100},
    width=.7\columnwidth,
    bar width=6mm,
    xlabel={Time in ms},
    yticklabel style={align=right},
    yticklabels={Tool setup, Ease of use, Tool documentation, Execution time, Format of input, Format of output},
    xmin=0,
    xmax=100,
    area legend,
    y=8mm,
    enlarge y limits={abs=0.625},
]
\addplot[findOptimalPartition,fill=findOptimalPartition] coordinates
{(6.7,0) (0,1) (13.3,2) (13.3,3) (6.7,4) (40,5) };
\addplot[storeClusterComponent,fill=storeClusterComponent] coordinates
{(33.3,0) (26.7,1) (46.7,2) (26.7,3) (60,4) (33.3,5)};
\addplot[dbscan,fill=dbscan] coordinates
{(20,0) (53.3,1) (26.7,2) (26.7,3) (20,4) (13.3,5)};
\addplot[constructCluster,fill=constructCluster] coordinates
{(40,0) (20,1) (13.3,2) (26.7,3) (6.7,4) (6.7,5)};
\addplot[gray!40,fill=gray!40] coordinates
{(0,0) (0,1) (0,2) (6.7,3) (6.7,4) (6.7,5)};
\legend{Very Satisfied, Somewhat Satisfied, Neutral, Somewhat Unsatisfied, Very Unsatisfied}
\end{axis}  
\end{tikzpicture}
\caption{Level of satisfaction with the aspects of SATDBailiff tool.}
\label{fig:likertscale}
\end{figure*}

 Figure \ref{fig:likertscale}  presents an overview of the aspects of the tool and the satisfaction of the participants. With respect to tool setup, some of the respondents reported that they are satisfied with the tool. However, there are a few participants who are usually not happy with how the tool is set up and we are planning on improving the tool setup in the future. For the easy to use aspect, a larger group (10 participants)  selected neutral. Some of the participants found that the tool is not easy to use, so we will work on improving the usability of the tool. The main feedback from the participants is to add GUI features, and we think that GUI might be an option to be considered. Adding GUI is not one of our priorities because we have more features that we would like to improve, but we will consider adding it in our future extension of the tool. Regarding the tool documentation, the majority of the respondents agreed that the documentation is useful; only 2 participants are somewhat unsatisfied. Concerning the execution time, the participants are happy with it, although we don’t have a consensus between participants and we believe that execution time depends on the selected project size. For the format of the input, the vast majority of respondents agree that the format is acceptable. Likewise, the participants are happy with the format of the output.
  
 With respect to the preferred features of the tools, the participants listed a variety of features which are centered around six main topics: (1) tool correctness, (2) tool accuracy, (3) tool documentation, (4) execution time, (5) data storage, and (6) Docker integration. We demonstrated some of the responses:
 
\begin{quote}
\par \say{\textit{The tool performed more smartly and accurately than I had anticipated, since one of my SATD resolving commits was actually in 2 parts because its was long, I assumed that would cause some inaccuracy and confusion in the tool. But surprising and delightfully the tool identified and clubbed the 2 commits and also connected to the reported SATD beautifully.}}

\par \say{\textit{The integration with Docker allowed the tool to be setup very easily on my system so I would have to say that 'addon' was fantastic. I found myself struggling to 1) Build the tool from the source code and 2) running the executable jar. Once I figured out how to download/install the image everything began to go smoothly for me. That integration is not necessarily a feature of the tool itself, so in that regard I would say I really liked the reports generated. The html + csv output were well designed and labels were intuitive. On top of that, I didn't have to do anything crazy to get the tool to run on my repo, just changed the repo url in the csv file which was quite simple. I appreciated that about the tool.}}

\par \say{\textit{I liked how fast are the SATD mining process and the correctness of the output}}

\par \say{\textit{I like how the tool is analyze the comments with high accuracy of detect each comment that was add or removed .Also , the way of the data is stored , the user have 3 options to analyze the data excel , html report and Sql database. In MySql database there are multiple tables with different information the user can create customize report based on his requirements.}}

\end{quote}
 
As for the suggested features of the tool, participants mainly mentioned 3 main features: (1) IDE integration, (2) issue troubleshooting , and (3) specifying the starting and ending commits. Participants did explicitly share their concerns during
the survey as follows:

\par \say{\textit{I feel that a frontend GUI that would help feeding in the data and also a simple GUI that did the MYSQL query in the backend and for us it would just be a download button would make the tool much more friendly and intuitive to use.” and “A GUI would be good as everything right now is terminal based.}}

\par \say{\textit{Add as a plugin and add more details about how to use the tool if a particular error is found.}}
 
\par \say{\textit{I think having a way to choose a starting and ending point would be great}}

\section{\approach Limitations and Upcoming Features}
\label{section:limitations}

In this section, we outline the known limitations to our tool, and the features we are planning on developing.

\begin{itemize}
    \item As pointed out in Section \ref{section:approach}, our current tool does not track comments introduced by commits with multiple parents. We have not yet performed any manual analysis to assess the extent to which this affects the tracking of current comments. In the future, we plan to perform a thorough examination to determine whether we should prioritize handling multiple parents.

\item To reduce the false positiveness of our tool, as pointed out by the experiments, we are currently exploring various text processing techniques to preprocess the comments and reduce the effects of special characters that may interfere with our string matching.

\item We are also investigating the use of pre-trained models that can detect whether comments are pointing out to the same technical debt. Such a model can help us avoid detecting two different comments as being the same one, in the case of removing one SATD and adding another one, for the same method, and the same commit.

\item Since the management of technical debt has been heavily correlated with refactoring, we have already integrated the Refactoring Miner library, through its API, to run on the commits that we are also detecting the removal of SATD in. This will support existing and future studies that analyze the necessary changes needed to remove technical debt. Our tool currently reports all the refactorings that are associated with the classes and methods containing SATD comments. This is a recent feature that we are still testing, therefore we do not have any observations about it in this paper. We anticipate that this feature will help researchers in developing empirical evidence of the usefulness of refactoring in terms of managing technical debt.

\item Upon performing the external validation, we found that the participants pointed out limitations that are mainly related to tool setup and usage. In the future, we plan to improve the setup and the usability of the tool by adding several features mentioned by the participants, including but not limited to,  IDE integration,  issue troubleshooting, and specifying the starting and ending commits. Further, the external validation is currently conducted with a group of software engineering students. As future work, we plan to perform another round of external validation with professional software engineers in industry to hear their perception.

\end{itemize}
\section{Threats to Validity}
\label{section:threats}

In this section, we identify potential threats to the validity of our approach and our experiments. 

Threats to the validity of this tool include the limited manual evaluation and general lack of testing. An important potential threat relates to our manual classification. Since the manual verification of samples is a human intensive task and it is subject to personal bias, we mitigate this first by selecting SATD instances from an existing dataset. Then we performed the tracking of their existence by one author. Only 200 samples of SATD were recorded and addressed to determine the accuracy of \approach. Ideally, this number should be much higher. There was limited time available for other formal and repeatable forms of programmatic testing, as a majority of the validation effort was allocated to the manual validation and validation done during development.

Another threat relates to the SATD instances that are extracted only from open source Java projects. Our results may not generalize to commercially developed projects, or to other projects using different programming languages. Another threat concerns the generalization of SATD patterns used in this study. Since a method is considered holder of TD when a comment contains SATD, this may not generalize to other projects if they do not allow inline documentation (comments). 

The reliance on the existing tool (i.e., SATD Detector) is another threat to validity due to the possibility of introducing false positive results. Thus, the binary classification of SATD comments resulting from the SATD Detector might have an impact on our findings.

\section{Conclusion and Future Work}
\label{section:conclusion}

This paper presents a preview of our tool, \approach , and discusses its benefits. The tool aims at offering an unmatched ability to extract SATD instances and the development operations upon them from Git repositories.
This paper discussed the benefits that a high quality empirical history would have for the further study of SATD.

An acceptable level of accuracy was achieved with \approach, however there is still opportunity for improvement. In their nature, source code differencing tools may not always record modifications that reflect the true nature of a developer's intentions \cite{Falleri}. It is because of these inaccuracies that solving each and every edge case is infeasible for the purpose of data generation. Some attempts to fix these edge-cases were made, however their large numbers and unpredictability made it a relatively futile task. To compensate, a list of known edge cases are included on the tool's website to detail known areas where inaccuracies may appear. As Edit Scripts are more reliably instantiated from these differences, more accurate detection of SATD-impacting operations will be possible.

Edit scripts were generated for this tool using the Histogram and Myers algorithms made available through JGit. Using a differencing tool like GumTree \cite{Falleri} or a hybrid approach  \cite{Matsumoto} may produce more accurate results. However, GumTree currently does not offer the tracking of comment changes and does not plan on implementing that functionality.\footnote{\url{https://github.com/GumTreeDiff/gumtree/issues/39}} Modification of the edit script generation methodology may also have positive impacts on the project's runtime which may currently be excessive for larger project.

Some other issues encountered include the handling of non-English language source comments. None of the projects used to validate \approach contained any known instances of these comments, however the larger dataset of 691 projects released alongside the tool does contain these instances. Ideally, a tool would be able to detect a non-English comment before classifying it as SATD, but no attempt was made to solve this issue as it only represents a small subset of the addressed projects.

\bibliographystyle{elsarticle-num}
\bibliography{paper}

\end{document}